\documentclass[reqno,a4paper]{amsart}
\usepackage{geometry, fancybox} 
\usepackage{graphicx}
\usepackage{tikz}
\usepackage{pgfplots}
\pgfplotsset{compat=1.3}
\usepackage{pdfsync}
\usepackage{xr}
\usepackage{color}
\usepackage{accents}
\usepackage{caption}
\usepackage{subcaption}
\usepackage{graphics}          
\usepackage{amsmath}
\usepackage{amsthm}
\newtheorem{thm}{Theorem}[section]

\newtheorem{definition}[thm]{Definition}

\newtheorem{remark}[thm]{Remark}

\usepackage{amssymb}
\usepackage{hyperref}
\usepackage[mathscr]{eucal}

\newcommand\Po{\textrm{P}_{\textrm{I}} }
\newcommand\Ptw{\textrm{P}_{\textrm{II}} }
\newcommand\Pth{\textrm{P}_{\textrm{34}} }
\newcommand\Pthree{\textrm{P}_{\textrm{III}}}

\numberwithin{equation}{section}
\numberwithin{figure}{section}
\newcommand\complex{\mathbb C}
\newcommand\real{\mathbb R}
\newcommand\proj{\mathbb P}

\begin{document}
\title[Geometric Asymptotic Analysis]{An Overview of Geometric Asymptotic Analysis\\ of Continuous and Discrete Painlev\'e Equations}
\author{Nalini Joshi}\thanks{The preparation of this paper was supported by an Australian Laureate Fellowship \# FL 120100094 and grant \# DP130100967 from the Australian Research Council.}
\address{School of Mathematics and Statistics F07, The University of Sydney, NSW 2006, Australia}
\email{nalini.joshi@sydney.edu.au}
\begin{abstract}
The classical Painlev\'e equations are so well known that it may come as a surprise to learn that the asymptotic description of its solutions remains incomplete. The problem lies mainly with the description of families of solutions in the complex domain. Where asymptotic descriptions are known, they are stated in the literature as valid for large connected domains, which include movable poles of families of solutions. However, asymptotic analysis necessarily assumes that the solutions are bounded and so these domains must be punctured at locations corresponding to movable poles, leading to asymptotic results that may not be uniformly valid. To overcome these issues, we recently carried out asymptotic analysis in Okamoto's geometric space of initial values for the first and second Painlev\'e equations. In this paper, we review this method  and indicate how it may be extended to the discrete Painlev\'e equations. 
\end{abstract}

\maketitle
\setcounter{tocdepth}{2}
\tableofcontents

\section{Introduction}
In a beautiful paper in 1979, Okamoto \cite{o:79} showed how to construct the space of initial values for each Painlev\'e equation, so that it includes initial values at a pole. Recently, we showed how to carry out asymptotic analysis in this space for the first \cite{dj:11} and second \cite{hj:13} Painlev\'e equations as the independent variable approaches infinity. We give an overview of the method in this paper and indicate how it may also extend to discrete Painlev\'e equations. 

The asymptotic analysis of the Painlev\'e equations started with Boutroux \cite{b:13} a hundred years ago.  Applications in statistical mechanics \cite{wmtb:76} and, more recently, in random matrix theory \cite{m:04,tw:94} revitalised modern interest in this area. Motivated by these applications, many different approaches have been developed for the asymptotic study of the Painlev\'e equations \cite{jk:88,jk:92,j:99,kt:05,k:94,dz:95,fikn:06}. However, known asymptotic results are valid only where the solutions and their derivatives are bounded. This constraint raises questions about the uniform validity of asymptotic behaviours for general solutions in large asymptotic domains, which may include movable poles as interior points. 

Given $t_0\in\complex$, which is a regular point for the Painlev\'e equation and bounded initial values $(y_0, y_1)$, the space of initial values is a manifold coordinatized locally by the solution $(y(t), y'(t))$, which satisfies $(y(t_0), y'(t_0))=(y_0, y_1)$.  Okamoto \cite{o:79} included unbounded initial values in this space by compactifying it and resolving any singularities. Following this approach, we recently investigated asymptotic analysis of the first \cite{dj:11} and second \cite{hj:13} Painlev\'e equations in initial value space and found new results on transcendental non-classical solutions in the complex plane.

In contrast to the classical Painlev\'e equations, very few asymptotic investigations have been carried out into discrete versions of the Painlev\'e equations, classified by Sakai \cite{s:01}. Three such studies are currently known: the additive difference first Painlev\'e equation \cite{j:97}, the $q$-discrete sixth Painlev\'e equation \cite{m:10}, and the $q$-discrete first Painlev\'e equation (q$\Po$) \cite{o:09,j:13}. Nishioka \cite{n:10} showed that q$\Po$, which has initial value space $A_7^{(1)}$, has no solutions expressible in terms of earlier known $q$-special functions. We have carried out analysis in the space of initial values for the latter equation and found some unusual new behaviours that arise in a neighbourhood of two coalescing base points in the asymptotic limit.

As the independent variable of the first five differential Painlev\'e equations approaches a limit, their general solutions approach elliptic functions to leading order (after a change of variables). Surprisingly, this does not appear to be the case for discrete Painlev\'e equations, when taking an asymptotic limit of the independent variable alone. Instead, a scaling limit that combines independent variable and parameters is needed to obtain elliptic-function-type leading-order behaviours. We illustrate this difference by using the $q$-discrete third Painlev\'e equation, q-$\Pthree$, as an example in Section \ref{say qp3}.

The plan of the paper is as follows. Local asymptotic limits of continuous and discrete Painlev\'e equations are described in Sections \ref{say} and \ref{say qp3}, with a particular focus on formative examples. In Section \ref{say}, the example is the first Painlev\'e equation in the limit as its independent variable approaches infinity. In Section \ref{say qp3}, the example is the third $q$-discrete Painlev\'e equation in two different limits, one in the independent variable alone and one in a scaling limit that combines independent variable and parameters. To explain how to analyse these further in initial value space, we introduce the explicit process of resolution of singularities for plane curves in Section \ref{plane}. In Section \ref{p1}, we consider asymptotic analysis of the first Painlev\'e equation within the geometric framework of initial-value space. In Section \ref{qp1}, we show how such a study extends to a $q$-discrete first Painlev\'e equation. We end the paper with a discussion in Section \ref{dis}.

\subsection{Background}\label{bg}
A search for new transcendental functions that generalise elliptic functions led Painlev\'e \cite{p:02,p:06} to the six ordinary differential equations that now bear his name. In this search, Painlev\'e was inspired by the definition of elliptic functions as solutions of differential equations. An alternative point of view is to consider them as solutions of difference equations, given by their addition formulas. This alternative view leads naturally to the discrete Painlev\'e equations.  

Elliptic functions arise again in the leading-order asymptotic behaviours of the Painlev\'e differential equations as the independent variable approaches a fixed singularity. To leading order in this limit, the first five Painlev\'e equations give rise to Hamiltonians that are invariant on curves of genus one. However, this is not the case for $q$-Painlev\'e equations.  When the independent variable approaches infinity (while parameters are held fixed) in a $q$-discrete Painlev\'e equation, we only get genus-zero curves. Approaching a genus-one curve requires more complicated scaling limits (involving a mixture of the independent variable and parameters) of such an equation. We provide an illustration of these results later in this paper through two main examples: $\Po$ in \S \ref{say} and q$\Pthree$ in \S \ref{say qp3}.

Given $g_2$, the Weierstrass elliptic function 
\begin{equation}\label{eq:weier_fn}
w(t)=\wp\bigl(t-t_0; g_2, g_3\bigl)
\end{equation}
satisfies
\begin{subequations}\label{eq:wp}
\begin{align}\label{eq:wp_ode}
w''&=6\,w^2-\,\frac{g_2}{2}\\
\label{eq:weier_integral}
w'^2&=4\,w^3-g_2\,w-g_3
\end{align}
\end{subequations}
where primes denote $t$-derivatives, for arbitrary parameters $t_0$ and $g_3$. Because $\wp(t; g_2, g_3)$ has poles at the origin, the general solution has poles located at $t_0$. This arbitrary constant is determined by initial values and so varies or \textit{moves} with different solutions. Such a singularity is called a \textit{movable} singularity. In this case, we see that all movable singularities are poles, due to the meromorphic nature of $\wp(t)$. If all movable singularities of all solutions of a differential equation are poles, the equation is said to have the Painlev\'e property. 

Aiming to generalise elliptic functions, Painlev\'e classified ODEs with this property by investigating all second-order ODEs in the class $y''=F(y', y, t)$, where $F$ is rational in $y'$ and $y$ and locally analytic in $t$. With contributions by Gambier \cite{g:09} and R. Fuchs \cite{f:05}, Painlev\'e's search culminated in six non-linear ODEs now called the Painlev\'e equations, whose general solutions are known to be new transcendental functions. The first of these is 
\begin{equation}\label{pone}
\Po : \quad w''=6\,w^2 + t, \quad w=w(t).
\end{equation}

Instead of starting with differential equations, the search for new functions could have started with difference equations. Such difference equations for elliptic functions arise from addition theorems, which provide maps on elliptic curves. Iteration of such maps on elliptic curves has a very long history, and corresponding maps were mentioned by Burnside in the context of finite order groups in 1911 \cite{bu:11}. 

The most general such maps are described through iteration on curves defined by biquadratic polynomials. 
Let $P(x, y)$ be a biquadratic polynomial, that is, a quadratic function of each variable. Then the level set $P(x, y)=0$ defines a biquadratic curve ${\mathcal P}$ and any given horizontal line or vertical line intersects ${\mathcal P}$ at two points. A QRT map \cite{qrt:88} is the composition of two actions on this curve: one switching the $x$-coordinates of two intersection points with a given horizontal line, and the other switching the $y$-coordinates of two intersections with a vertical line \cite{t:04}. 

The map is, therefore, defined by the composition of two involutions, which each preserve the same biquadratic invariant
\[
{\displaystyle I(x,y)=\frac{{\bf X}^TA_0{\bf
Y}}{{\bf X}^TA_1{\bf Y}}},
\]
where ${\bf X}$, ${\bf Y}$ are vectors ${\bf X}=(x^2,x,1)^T,$ ${\bf Y}=(y^2,y,1)^T$ and $A_0,$ $A_1$ are  $3\times 3$ matrices. The general QRT map is the composition $\phi:=\iota_2\circ \iota_1$ of the involutions $\iota_1,$ $\iota_2,$ defined by the solutions of the equations $I(\hat x,y)-I(x,y)=0$ and $I(x,\widetilde y)-I(x,y)=0$ respectively:
\begin{equation} \label{QRT}
\iota_1:\left\{ \begin{array}{l}
\hat x={\displaystyle \frac{f_1(y)-f_2(y)x}{f_2(y)-f_3(y)x}}\\[3mm]
  \hat y=y
    \end{array}\right.,\quad
\iota_2:\left\{\begin{array}{l}
\widetilde x=x\\
  \widetilde y={\displaystyle\frac{g_1(x)-g_2(x)y}{g_2(x)-g_3(x)y}}
    \end{array}\right.
    \end{equation}
    where
    \begin{equation}
\begin{array}{c}
\bigl(f_1(y),f_2(y),f_3(y)\bigr)^T     = (A_0{\bf Y})\times(A_1{\bf Y}),\\
\bigl(g_1(x),g_2(x),g_3(x)\bigr)^T = (A_0^T{\bf X})\times(A_1^T{\bf X}).
    \end{array}
\end{equation}
The map $\phi$ bi-rationally preserves the linear pencil of bi-quadratic curves
$$
B(x,y;t):={\bf X}^TA_0{\bf Y}-I\;{\bf X}^TA_1{\bf Y}.
$$
In the symmetric case, when $A_i=A_i^T$, the map is defined by the composition of either $\iota_1$ or $\iota_2$ with the involution $j:\ x\mapsto y,\ y \mapsto x$, which leads to the three-point map
\begin{equation}\label{symm_qrt}
\overline y = \frac{f_1(y)-f_2(y)\underline y}{f_2(y)-f_3(y)\underline y}.
\end{equation}
In this case, the biquadratic pencil is parameterized by Jacobian elliptic functions (see, e.g., \cite[\S 15.10]{b:82}).

The generalisation of such maps also has a long history. The first example goes back to the work of Shohat \cite{s:39} in 1939 on orthogonal polynomials. More recently, non-linear non-autonomous difference equations were found as mappings arising from Schlesinger transformations in the theory of  monodromy preserving deformations of linear ODEs by Jimbo and Miwa \cite{jm:81} and as exactly solvable cases in string theory by Periwal and Shevitz \cite{ps:90}. Many subsequent examples were deduced by Grammaticos, Ramani and Hietarinta by extending the QRT map to equations possessing the singularity confinement property, at the time regarded as a discrete version of the Painlev\'e property \cite{rgh:91}. For each continuous Painlev\'e equation, there are now known to be many integrable second-order discrete versions. 

By showing how to obtain discrete Painlev\'e equations as mappings on rational surfaces obtained from a nine-point blow-up of the complex projective plane, Sakai \cite{s:01} resolved and classified these equations. One of these is the symmetric discrete third Painlev\'e equation, which has the form
\begin{equation}\label{dp3}
\overline w\,\underline w=\gamma\,\delta\,\frac{(w -\,\alpha\,t)\,(w-\,\beta\,t)}{(w -\,\gamma)\,(w-\,\delta)}
\end{equation}
where $t=t_0\,q^n$ and $q\not=0, 1$, $\alpha$, $\beta$, $\gamma$, $\delta$, $t_0$ are constant parameters, and we use the abbreviation $\overline w=w(q\,t)$, $\underline w=w(t/q)$. A degeneration \cite{j:13} of this equation leads to the $q$-discrete first Painlev\'e equation 
\begin{equation}\label{eq: qp1 scaled}
{\rm q}\Po :\quad  \overline w\,\underline w =\frac{t\,w-\,1}{t\,w^2}
\end{equation}

\section{Local asymptotic analysis of the Painlev\'e equations}\label{say}
To analyse the asymptotic behaviour of solutions of ordinary differential equations as the independent variable approaches infinity, it is useful to find an analytic change of variables in which the sizes of various terms in the equation become more explicit. In this section, we consider the first Painlev\'e equation as the independent variable approaches infinity. 

Taking $w_1=w$, $w_2=w'$, the first Painlev\'e equation \eqref{pone} in system form  is
\begin{equation}\label{p1: system}
\frac{d}{dt}\begin{pmatrix}w_1\\ w_2\end{pmatrix}
  =\begin{pmatrix}w_2\\ 6\,w_1^2+t \end{pmatrix}
\end{equation}
It has a $t$-dependent Hamiltonian
\begin{equation}\label{p1_Hamiltonian}
H=\frac{w_2^2}{2}-2\,w_1^3-t\,w_1.
\end{equation}
That is, Equations \eqref{p1: system} are equivalent to Hamiltonian's equations of motion: $dw_1/dt= \partial H/\partial w_2$, $dw_2/dt=-\,\partial H/\partial w_1$.

To study the solutions of $\Po$ as $|t|\to\infty$, we transform it to Boutroux's coordinates
\begin{subequations}\label{p1: boutroux trans}
\begin{eqnarray}
w_1&=&t^{1/2}\,u_1(z)\\
w_2&=&t^{3/4}\,u_2(z)\\
z&=&\frac{4}{5}\,t^{5/4}
\end{eqnarray}
\end{subequations}
Such a transformation of variables can be found by assuming $w_j(t)=f_j(t)\,u_j(z)$, $z=g(t)$, then deducing $f_j$, $g$ from possible balances of dominant terms of the transformed system in the limit. 

The transformation \eqref{p1: boutroux trans} converts \eqref{p1: system} to 
\begin{equation}\label{p1: boutroux system}
\frac{d}{dt}\begin{pmatrix}u_1\\ u_2\end{pmatrix}
  =\begin{pmatrix}u_2\\ 6\,u_1^2+1 \end{pmatrix}
      -\,\frac{1}{(5z)}\begin{pmatrix}2\,u_1\\ 3\,u_2 \end{pmatrix}
\end{equation}
The Hamiltonian $H$ is now replaced by 
\begin{equation}\label{p1: E}
E:=\frac{u_2^2}{2}-2\,u_1^3-u_1 .
\end{equation}
An explicit calculation shows that 
\begin{equation}\label{p1: E_z}
\frac{dE}{dz}=-\,\frac{1}{(5z)}\,\left(6\,E+4\,u_1\right).
\end{equation}
As $|z|\to\infty$, along any path where $u_1$ stays bounded, $E$ approaches a constant value, which may not be a uniform limit for all paths of approach to infinity. Moreover, the leading-order asymptotic behaviour is given by $u_1\sim\,\wp(t-t_0; -12, -\,2\,E)$, where $t_0$ is arbitrary, along such a path. 

So the values of $E$ provide level curves of the cubic function
\begin{equation}\label{eq:cubic}
f(x, y)=y^2 - 4\,x^3+g_2\,x
\end{equation}
for $g_2=-2$. The level curves $f=-g_3$ for the case $g_2=12$ with $g_3\le0$ are illustrated in Figure \ref{fish-eye}. The slow evolution of $E$ implies that $g_3$ drifts slowly as $t$ approaches infinity. In other words, Figure \ref{fish-eye} is a snapshot of possible curves on which the solution moves, with each curve changing slowly as $t$ changes in this limit.
 \pgfplotsset{no markers}
\begin{figure}
\begin{tikzpicture}[auto,smooth,samples=120,domain=0:3,xscale=1.4,yscale=0.88]
  \draw[->] (-3.9,0) -- (3.9,0) node[right] {$x$};
  \draw[->] (0,-4.2) -- (0,4.2) node[above] {$y$};
		{
  	\draw[color=red]   plot [domain=-2.0:1] (\x, {2*sqrt( (\x + 2) )*(\x -1) })  ;
  	\draw[color=red]   plot [domain=-2.0:1] (\x, {- 2*sqrt( (\x + 2) )*(\x -1) })  ;
  	\draw[color=red]   plot [domain=1.0:2.2] (\x, {2*sqrt( (\x + 2) )*(\x -1) })  ;
  	\draw[color=red]   plot [domain=1.0:2.2] (\x, {- 2*sqrt( (\x + 2) )*(\x -1) })  ;
  	\draw[color=blue]   plot [domain=-2.02727:2.2] (\x, {sqrt( 4*\x^3-12*\x+9) })  ;
  	\draw[color=blue]   plot [domain=-2.02727:2.2] (\x, {- sqrt( 4*\x^3-12*\x+9) })  ;
  	\draw[color=blue]   plot [domain=-2.079106:2.2] (\x, {sqrt( 4*\x^3-12*\x+11) })  ;
  	\draw[color=blue]   plot [domain=-2.079106:2.2] (\x, {- sqrt( 4*\x^3-12*\x+11) })  ;
  	\draw[color=blue]   plot [domain=-2.1277:2.2] (\x, {sqrt( 4*\x^3-12*\x+13) })  ;
  	\draw[color=blue]   plot [domain=-2.1277:2.2] (\x, {- sqrt( 4*\x^3-12*\x+13) })  ;
  	\draw[color=blue]   plot [domain=-1.9422:0.55787] (\x, {sqrt( 4*\x^3-12*\x+6) })  ;
  	\draw[color=blue]   plot [domain=-1.9422:0.55787] (\x, {- sqrt( 4*\x^3-12*\x+6) })  ;
  	\draw[color=blue]   plot [domain=1.3844:2.2] (\x, {sqrt( 4*\x^3-12*\x+6) })  ;
  	\draw[color=blue]   plot [domain=1.3844:2.2] (\x, {- sqrt( 4*\x^3-12*\x+6) })  ;
  	\draw[color=blue]   plot [domain=-1.9115:0.4463] (\x, {sqrt( 4*\x^3-12*\x+5) })  ;
  	\draw[color=blue]   plot [domain=-1.9115:0.4463] (\x, {- sqrt( 4*\x^3-12*\x+5) })  ;
  	\draw[color=blue]   plot [domain=1.46524:2.2] (\x, {sqrt( 4*\x^3-12*\x+5) })  ;
  	\draw[color=blue]   plot [domain=1.46524:2.2] (\x, {- sqrt( 4*\x^3-12*\x+5) })  ;
		}
  \end{tikzpicture}\caption{Level curves of $f=y^2 - 4\,x^3+12\,x=-g_3$ for $-13\le g_3\le -5$. The curve drawn in red is the case $g_3=-8$ in which two of the zeroes of $f(x,y)$ merge at $x=1$.} \label{fish-eye}
\end{figure}
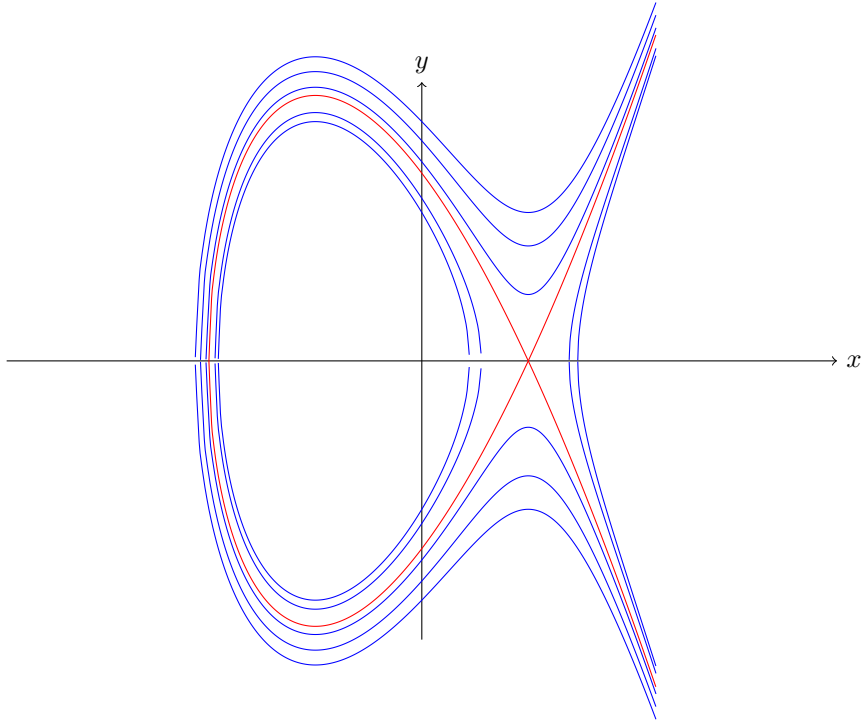

For the second Painlev\'e equation in system form 
\begin{equation}\label{p2: system}
\frac{d}{dt}\begin{pmatrix}w_1\\ w_2\end{pmatrix}
  =\begin{pmatrix}w_2\\ 2\,w_1^3+t\,w_1+\alpha \end{pmatrix}
\end{equation}
Boutroux's change of variables is $z=2t^{3/2}/3$, with $w_1=t^{1/2}\,u(z)$. To produce a more symmetric system, we take $w_2=t\,(v(z)-u(z)^2-1/2)$, which leads to  
 \begin{eqnarray}\label{p2: boutrouxsys}
\dot{u} & = & v-u^2-1/2-\dfrac{u}{3z},  \\ \label{boutrouxsys2}
\dot{v} & = & 2uv+\dfrac{2\alpha+1}{3z}-\dfrac{2v}{3z} .
\end{eqnarray}
This system allows us to treat both $\Ptw$ and $\Pth$ (which are related through a Miura transformation) simultaneously, because here $u$ and $v$ satisfy the Boutroux forms of $\Ptw$ and $\Pth$ respectively:
\begin{subequations}
\begin{align}
\label{bp2}
 \ddot{u}&=2u^3+u+\dfrac{2\alpha}{3z}-\dot{u}\dfrac{1}{z}+\dfrac{u}{9z^2} \\ 
  \ddot{v}&=\dfrac{\dot{v}^2}{2v}+2v^2-v-\dfrac{\dot{v}}{z}-\dfrac{(2\alpha-1)^2}{18vz^2}+\dfrac{1}{4z^2}.  \label{bp32}  
\end{align}
\end{subequations}
The Boutroux system (\ref{p2: boutrouxsys}) is an order $z^{-1}$ perturbation of the autonomous system with time independent Hamiltonian given by
\begin{equation}\label{energyfct}
E:=v^2/2-u^2v-v/2,
\end{equation}
where 
\begin{equation}
\dot{E}:=-\,\dfrac{4E}{3z}+\dfrac{4\alpha v-(2\alpha+1)(2u^2+1)}{6z}\,\,.
\end{equation}
Now the solutions parametrize slowly varying bi-quadratic curves in the $(u, v)$-plane.

\section{Local asymptotic analysis of discrete Painlev\'e equations}\label{say qp3}
To study the asymptotic behaviour of solutions of difference equations in limits, we again look for a change of variables in which the sizes of different terms in the equation become more explicit. In this section, we consider the $q$-discrete third Painlev\'e equation in two limits, firstly as the independent variable approaches infinity with parameters fixed and secondly as both independent variable and parameters approach a scaled limit. 

Following Boutroux \cite{b:13}, we rescale $w(t)=f(t)\,u(t)$ in order to make the asymptotic behaviours admitted by Equation \eqref{dp3}, as $t\to\infty$, more explicit. (To be specific, we also assume that $|q|>1$ and $0<\Re n$, with $n\gg 1$.) Assuming $u={\mathcal O}(1)$, we find  only one case that is self-consistent:
\begin{align}
f(t)={\mathcal O}(t^p), 0\le p<1\ &\Rightarrow\ \overline u\,\underline u\, \overline f\,\underline f\sim \gamma\,\delta\,\frac{\alpha\,\beta\,t^2}{u^2\,f^2}
\end{align}
We deduce that the only consistent scaling must satisfy $2p=2-2p$, which implies $f(t)=t^{1/2}$. This scaling results in the leading order equation
\begin{equation}\label{leading dp30}
\overline u\,\underline u=\frac{\alpha\,\beta\,\gamma\,\delta}{u^2}.
\end{equation}
This autonomous equation has an invariant
\begin{equation}\label{ap3}
I(x, y) =\frac{x^2\,y^2+\alpha\,\beta\,\gamma\,\delta}{x\,y}.
\end{equation}
Solving the polynomial equation $x^2\,y^2-K\,x\,y+\rho=0$, where $\rho=\alpha\,\beta\,\gamma\,\delta$, $I(x,y)=K$, we find
\[
x\,y=\frac{K}{2}\pm\sqrt{\frac{K^2}{4}-\rho}
\]
Therefore, the leading-order invariant curves are not elliptic curves, but conics (actually hyperbol\ae ). In the same way, as $t\to\infty$, we get the leading-order limit of q$\Po$ 
\begin{equation}\label{eq: qp1 auto}
 \overline{\mathtt{w}}\,{\mathtt{w}}\,\underline{\mathtt{w}} =1
\end{equation}
which has an invariant given by
\begin{equation}\label{ap1}
I(x, y) =\frac{x^2\,y^2+x\,y+x+y}{x\,y}
\end{equation}
i.e., $I(\overline{\mathtt{w}}, \mathtt{w})-I({\mathtt{w}}, \underline{\mathtt{w}})=0$ when $\mathtt{w}$ satisfies Equation \eqref{eq: qp1 auto}. The only finite points where the gradient of $I(x, y)$ vanishes is given by $x=y=\omega$ where $\omega^3=1$. 

To get more general leading-order behaviours, we investigate a scaling limit in which $t$ approaches infinity at the same time as some of the parameters. Scaling $t=\eta\,(1+z/\eta)$, $\alpha=\eta^{-1}\,a$, $\beta=\eta^{-1}\,b$, gives to leading order
\begin{equation}\label{ell-p3}
\overline w\,\underline w=\gamma\,\delta\,\frac{(w -\,a)\,(w-\,b)}{(w -\,\gamma)\,(w-\,\delta)}
\end{equation}
which is a symmetric QRT map with invariant
\[
I(x, y)=\frac{x^2\,y^2 - x\,y\,(\delta + \gamma)\,(x+y) + (x^2+y^2)\,\delta\,\gamma -\,(x+y)\,\delta\,\gamma\,(a + b) + a\,b\,\delta\,\gamma}{x\,y}
\]
This is parametrized by elliptic functions (for generic values of $a$, $b$, $\gamma$, $\delta$). 

However, the domain of asymptotic validity of this behaviour is constrained by $z\ll \eta$. Assuming $t_0=1$ without loss of generality (by rescaling $\eta$ if necessary), this constraint means the asymptotic analysis is valid only in a bounded local patch in $n$ near infinity given by $n\sim (\log\eta+z/\eta)/\log q$ for sufficiently large $\eta$.  Extending this domain to arbitrarily large $n$ will necessarily impose a process of degeneration on the leading-order elliptic behaviours, leading to the behaviours on the genus zero curve given by Equation \eqref{ap3}. Note also that imposing the requirement of bounded iteration in $n$ and $z$ leads to the condition $q\to1$ as $\eta\to\infty$. 

\section{Plane Algebraic Curves Arising in Limits}\label{plane}
To understand how to describe asymptotic behaviours in initial value space, we need an explicit description of the process of resolution of singularities.  We provide a short introduction here for interested readers. More information may be found in standard texts on complex algebraic curves \cite{g:89,k:92,w:04}.

For given $g_2$, and arbitrary $g_3$, the level curves \eqref{eq:cubic} form a family of curves in which each curve is identified by a distinct value of $g_3$. This \textit{pencil} of curves forms the integral curves of a vector field given by the Boutroux-Painlev\'e system \eqref{p1: boutroux system}. Recalling that $g_3=-\,2E$ varies slowly with time (see \eqref{p1: E_z}), each integral curve will change slowly within the pencil as time evolves. 

Given bounded initial values at a point in time, a solution of Equations \eqref{p1: boutroux system} exists locally and is uniquely defined by standard theorems of the theory of ODEs. To include unbounded initial values, we need to work in homogeneous coordinates $[x: y:1]=[u/w: v/w: 1]=[u: v: w]$ (for $[u: v: w]\not=[0: 0:0]$), which define a point in $\proj^2 (=\complex\proj^2)$. In such coordinates, the equation defining the pencil becomes 
\[
F(u, v, w)=w\,v^2-4\,u^3+g_2\,u\,w^2+g_3\,w^3=0.
\]
Notice that these curves all intersect (for arbitrary $g_3$) when $w=0$ and $u=0$, i.e., at $b_0=[0:1:0]$ on the line at infinity $L_0$ in $\proj^2$. Such an intersection point is called a \textit{base point} of the pencil. The value of $g_3$ is left free at such a point and, therefore, information about how to uniquely continue a solution through the point is lost. 

In a similar way, the functions that parametrize an integral curve $f(x, y)=0$ of a vector field $\dot x=P(x, y)$, $\dot y=Q(x, y)$ also become ill-defined at points where $\nabla f=0$. Such points are called \lq\lq singularities\rq\rq\ of the curve, because the tangent to the curve is no longer defined uniquely at such points. An example is given by $f(x, y)=y^2-4\,x^3=0$ (i.e., the cubic curve \eqref{eq:cubic} when $g_2=0$). This curve is singular at $b_0=(0,0)$, since $\nabla f = (2 y, -\,12\,x^2)$ vanishes at the origin, and has a cusp there. A graph of the curve is drawn in Figure \ref{cubic cusp}. 

From the perspective of differential equations theory,  both types of points, i.e., base points of pencils of curves and singularities of a curve, are problematic because solutions fail to be defined uniquely through such points. However, there is a way of resolving this ambiguity by using the tangential directions of the curves to distinguish them as they approach the problematic point. The process of \lq\lq blowing-up\rq\rq\ a point in algebraic geometry uses this key idea, which was due originally to Newton.

We illustrate the explicit process of blowing up here for the singularity at the origin of the curve $f(x, y)=y^2-4\,x^3=0$. In each case, notice that the new coordinates involve the ratio of the old coordinates, i.e., the slope of the tangent of the curve as it approaches the singularity is used as one of the new coordinates.
\pgfplotsset{no markers}
\begin{figure}\centering
\begin{subfigure}{0.4\textwidth}\centering	
	\begin{tikzpicture}[auto,domain=0:2]
  	 	\draw[->] (-2.2,0) -- (2.2,0) node[right] {$x$};
 		 \draw[->] (0,-2.2) -- (0,2.2) node[above] {$y$};
 		 \draw[color=blue]   plot [domain=0:1] (\x, {sqrt( (4*\x^3) ) }) node[right] {$\mathcal C$}  ;
 		 \draw[color=blue]   plot [domain=0:1] (\x, {-sqrt( (4*\x^3) ) })  ;
 	 \end{tikzpicture}
	 \caption{Graph of $f$: $y^2=4\,x^3$} \label{cubic cusp}
\end{subfigure}
\begin{subfigure}{0.4\textwidth}\centering	
	 \begin{tikzpicture}[auto,domain=0:2]
 		 \draw[->] (-2.2,0) -- (2.2,0) node[right] {$x_1$};
 		 \draw[blue,->] (0,-2.2) -- (0,2.2) node[black,above] {$y_1$};
  		\draw[color=blue]   plot [domain=0:1] (\x, {sqrt( 4*(\x) ) }) node[right] {${\mathcal C}^{(1)}$}  ;
  		\draw[color=blue]   plot [domain=0:1] (\x, {-sqrt( 4*(\x) ) })  ;
  		\coordinate [label=below:{${\mathcal E}_0$}]  (A) at (-0.4,1.8)  ;
	\end{tikzpicture}
  	\caption{The total transform of $f$} \label{strict f1}
\end{subfigure}
\begin{subfigure}{0.4\textwidth}\centering	
	\begin{tikzpicture}[auto,domain=0:2]
  		\draw[blue,->] (-2.2,0) -- (2.2,0) node[black,right] {$x_2$};
  		\draw[blue,->] (0,-2.2) -- (0,2.2) node[black,above] {$y_2$};
 		 \draw[color=blue]   plot [domain=-0.5:0.5] (\x, {4* \x}) node[right] {${\mathcal C}^{(2)}$}  ;
 		 \coordinate [label=below:{${\mathcal E}_0^{(1)}$}]  (A) at (-0.4,+2)  ;
  		 \coordinate [label=right:{${\mathcal E}_1$}]  (B) at (+2,-0.4)  ;
   \end{tikzpicture}
   \caption{Total Transform of $f_1$}\label{strict f2}
\end{subfigure}
\begin{subfigure}{0.4\textwidth}\centering	
   	\begin{tikzpicture}[auto,domain=0:2]
  		\draw[blue,->] (-2.4,0) -- (2.4,0) node[black,right] {$x_3$};
  		\draw[blue,->] (0,-2.2) -- (0,2.2) node[black,above] {$y_3$};
 		 \draw[blue,dashed] (-1.8,-1.8) -- (-1.8,1.8) node[black,left] {${\mathcal E}_0^{(2)}$};
 		 \draw[color=blue]   plot [domain=-2.2:2.2] (\x, 1) node[right] {${\mathcal C}^{(3)}$}  ;
   		\coordinate [label=below:{${\mathcal E}_1^{(1)}$}]  (B) at (-0.4,2)  ;
   		\coordinate [label=below:{${\mathcal E}_2$}]  (C) at (+2,-0.2)  ;
   \end{tikzpicture}
   \caption{Total Transform of $f_2$}   \label{strict f3}
  \end{subfigure}
\caption{Resolution of the singularity at the origin on the curve $f(x, y)=y^2-4\,x^3=0$.}
\label{res cubic}
\end{figure}
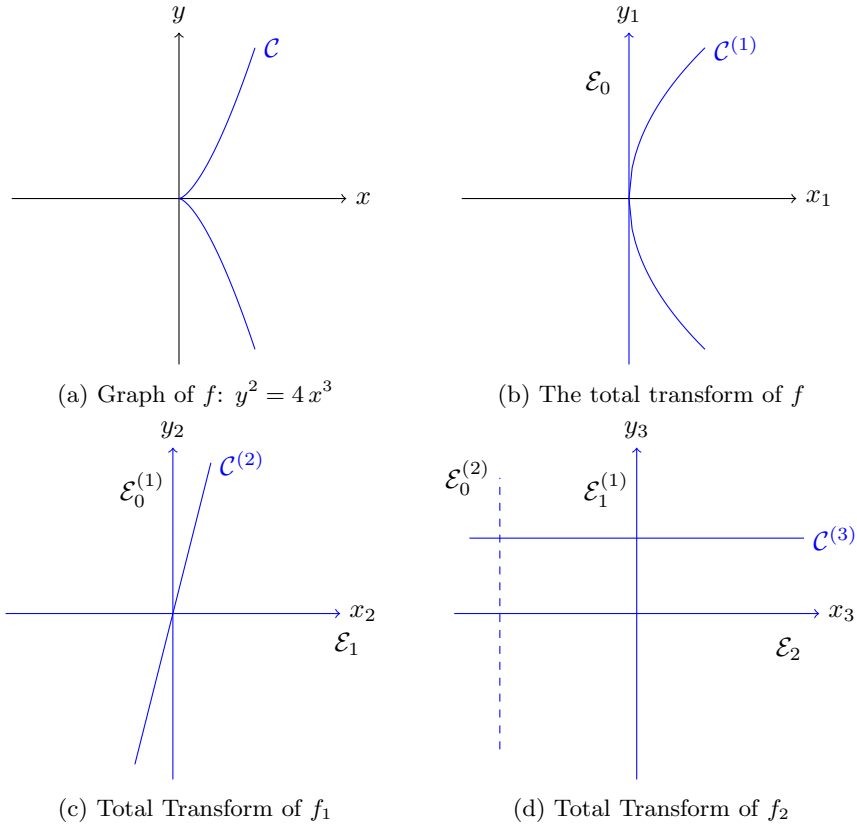
To resolve this point, the blow-up of the origin is achieved by changing variables to 
\[
(x, y)=(x_1, x_1\,y_1)\quad\Rightarrow\quad x_1=x, y_1={y}/{x}.
\]
Then $f$ becomes
\begin{equation}\label{x1}
f(x, y)=x_1^2\,y_1^2-\,4\,x_1^3=x_1^2\,(y_1^2-\,4\,x_1).
\end{equation}
This is called \lq\lq total transform\rq\rq\ of $f$.  The factorisation in Equation \eqref{x1} transforms the curve to the product of a line $x_1=0$ and a parabola $y_1^2-4\,x_1=0$, as illustrated in Figure \ref{strict f1}.  
The line $x_1=0$ is called the \lq\lq exceptional line\rq\rq\ ${\mathcal E}_0$ (counted twice). The second factor $f_1:=y_1^2-4\,x_1$ is called the \lq\lq strict transform\rq\rq\ of $f$ and the corresponding equation $f_1=0$ provides the strict transform ${\mathcal C}^{(1)}$ of the original curve ${\mathcal C}$.

Note that $f_1$ is no longer singular and that the curve ${\mathcal C}^{(1)}$ is smooth. However, this is not good enough because ${\mathcal C}^{(1)}$ touches ${\mathcal E}_0$ tangentially (at the origin). So there is still an ambiguity with continuing a solution on the curve through the origin. In other words, we have to blow up again. We do so by transforming coordinates to 
\[
(x_1, y_1)=(x_2\,y_2, y_2)\quad\Rightarrow\quad x_2={x_1}/{y_1}, y_2=y_1,
\]
which gives
\begin{equation}\label{x2}
f_1(x_1, y_1)=y_2^2-\,4\,x_2\,y_2=y_2\,(y_2-\,4\,x_2).
\end{equation}
There is a new exceptional line ${\mathcal E}_1$ given by $y_2=0$ and the strict transform of $f_1$ is $f_2:=y_2-\,4\,x_2$. Note that in this chart, the \lq\lq lift\rq\rq\ of ${\mathcal E}_0$, which is denoted ${\mathcal E}_0^{(1)}$, is still visible and given by $x_2=0$. The strict transform of the curve ${\mathcal C}^{(2)}$, given by $f_2=0$,  is no longer tangential to the exceptional lines, however, it intersects both ${\mathcal E}_0^{(1)}$ and ${\mathcal E}_1$ simultaneously at the origin and so the flow is still problematic there. 
We blow up again by transforming coordinates to 
\[
(x_2, y_2)=(x_3, x_3\,y_3)\quad\Rightarrow\quad x_3=x_2, y_3={y_2}/{x_2},
\]
which gives
\begin{equation}\label{x2}
f_2(x_2, y_2)=x_3\,y_3-\,4\,x_3=x_3\,(y_3-\,4).
\end{equation}
Now we have a new exceptional line ${\mathcal E}_2$ given by $x_3=0$ and the transform of the curve ${\mathcal C}^{(3)}$ is the horizontal line $y_3=4$. The lift of the exceptional line ${\mathcal E}_1^{(1)}$ is still visible and is given by $y_3=0$. However, ${\mathcal E}_0^{(2)}$ is no longer visible as it now lies in the chart where $x_3$ is infinite. All the exceptional lines and the curve now intersect each other transversally at distinct points, as shown in Figure \ref{strict f3}. This result is called a \textit{good resolution} and provides a space in which the functions parametrizing the curve are uniquely defined every where.

This process can also be applied to base points in $\proj^2$. A good resolution is guaranteed by standard theorems in algebraic geometry \cite{g:89,w:04}.
\section{The first Painlev\'e equation in the limit $t\to\infty$ in initial value space}\label{p1}
To investigate  the asymptotic behaviours of $\Po$ in domains that include poles of the solutions, we embed the Boutroux-Painlev\'e system \eqref{p1: boutroux system} in $\proj^2$ by identifying the affine coordinates $(u_{1}, u_{2})$ with homogeneous coordinates as follows
\begin{eqnarray*}
 [1:u_{1}:u_{2}]&=&[u_1^{-1}:1:u_1^{-1}\,u_2]=:[u_{021}:1:u_{022}]\\
 &=&[u_2^{-1}:u_1\,u_2^{-1}:1]=:[u_{031}:u_{032}:1]
 \end{eqnarray*}
The line at infinity given respectively by $u_{021}=0$, $u_{031}=0$ is denoted by ${\mathcal E}_0$.  For $0\le i \le 8$, corresponding to the $i$-th stage of the regularizing (blow-up) sequence, we denote a base point by $b_{i}$, the exceptional line attached to that base point by ${\mathcal E}_{i+1}$ and the coordinates in the $j$-th chart of the $i$-th blowup by $(u_{ij1}, u_{ij2})$. At each stage of the blow-up process, we encounter two charts, which are labelled by $j=1, 2$. In each coordinate chart, the Jacobian of the coordinate change from $(u_1, u_2)$ to $(u_{ij1}, u_{ij2})$ is denoted by
\begin{equation*}
 w_{ij}=\frac{\partial u_{ij1}}{\partial u_1}\,\frac{\partial u_{ij2}}{\partial u_2}-\frac{\partial u_{ij1}}{\partial u_2}\,\frac{\partial u_{ij2}}{\partial u_1}.
\end{equation*}
We refer the reader to the paper \cite{dj:11} for full and explicit details of the process of resolution. Nine blow-ups are needed, and the sequence of base points is given by
\begin{align}
\begin{split}
(u_{031},u_{032})&\xleftarrow{b_0=(0,0)} (u_{111},u_{112})\xleftarrow{b_1=(0,0)} (u_{211},u_{212})\xleftarrow{b_2=(0,0)} (u_{311},u_{312})     \\
\label{seq}
   &\xleftarrow{b_3=(4,0)} (u_{411},u_{412}) \xleftarrow{b_4=(0,0)}  (u_{511},u_{512}) \xleftarrow{b_5=(0,0)}  (u_{611},u_{612})\\
 &   
   \xleftarrow{b_6=(0,0)} (u_{711},u_{712}) \xleftarrow{b_7=(32,0)} (u_{811},u_{812}) \xleftarrow{b_8=(-256/(5z),0)}  (u_{911},v_{912})    \end{split}
\end{align}
Note that the first eight are identical to those for the pencil of cubic curves: $y^2-4\,x^3 + 2\,x+g_3=0$, whilst the ninth one differs from the last base point of this pencil by order ${\mathcal O}(1/z)$. The intersection diagram of the result is given in Figure \ref{e8}. For each $z$, the resolved space we construct is denoted by ${\mathcal S}_9(z)$. 

\pgfplotsset{no markers}
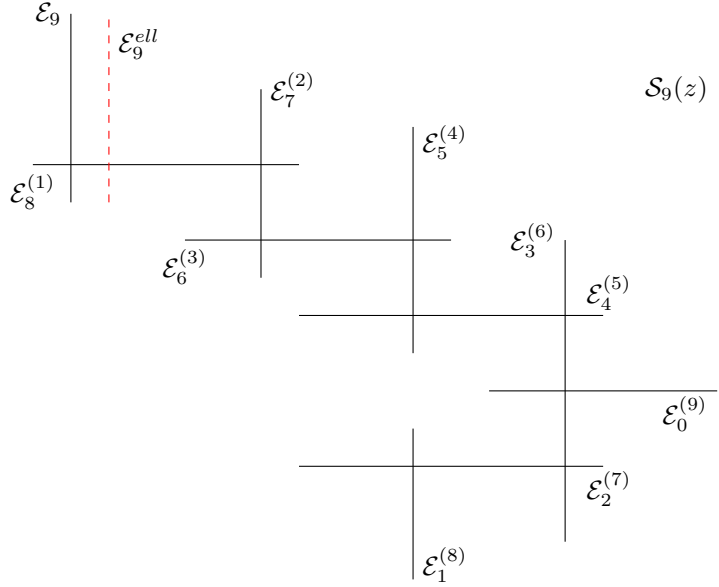
\begin{figure}
\begin{tikzpicture}[auto,smooth,domain=0:6]
  		\draw[-] (1,0) -- (4,0) ; \coordinate [label=left:{${\mathcal E}_0^{(9)}$}]  (e0) at (4,-0.3)  ;
  		\draw[-] (0,-2.5) -- (0,-0.5) ; \coordinate [label=right:{${\mathcal E}_1^{(8)}$}]  (e1) at (0,-2.3)  ;
  		\draw[-] (-1.5,-1) -- (2.5,-1) ; \coordinate [label=left:{${\mathcal E}_2^{(7)}$}]  (e2) at (3,-1.3)  ;
  		\draw[-] (2,-2) -- (2,2) ; \coordinate [label=left:{${\mathcal E}_3^{(6)}$}]  (e3) at (2,2)  ;
  		\draw[-] (-1.5,1) -- (2.5,1) ; \coordinate [label=left:{${\mathcal E}_4^{(5)}$}]  (e4) at (3,1.3)  ;
  		\draw[-] (0,0.5) -- (0,3.5) ; \coordinate [label=right:{${\mathcal E}_5^{(4)}$}]  (e5) at (0,3.3)  ;
  		\draw[-] (-3,2) -- (0.5,2) ; \coordinate [label=below:{${\mathcal E}_6^{(3)}$}]  (e6) at (-3,2)  ;
   		\draw[-] (-2,1.5) -- (-2,4) ; \coordinate [label=right:{${\mathcal E}_7^{(2)}$}]  (e7) at (-2,4)  ;
  		\draw[-] (-5,3) -- (-1.5,3) ; \coordinate [label=below:{${\mathcal E}_8^{(1)}$}]  (e8) at (-5,3)  ;
   		\draw[-] (-4.5,2.5) -- (-4.5,5) ; \coordinate [label=left:{${\mathcal E}_9$}]  (e9) at (-4.5,5)  ;
   		\draw[red,dashed,-] (-4,2.5) -- (-4,5) ; \coordinate [label=right:{${\mathcal E}^{ell}_9$}]  (e90) at (-4,4.6)  ;
   		 \coordinate [label=left:{$\mathcal S_9(z)$}]  (s9) at (4,4)  ;
 \end{tikzpicture}\caption{Intersection diagram of exceptional lines for $\Po$. For each $z$, Okamoto's space is this resolved space $\mathcal S_9(z)$ with the union of exceptional lines ${\mathcal I}=U_{j=0}^8\,{\mathcal E}_j^{(9-j)}$ removed. The dashed line in red indicates the ninth exceptional line ${\mathcal E}^{ell}_9$ for the elliptic case, which is ${\mathcal O}(1/z)$ close to ${\mathcal E}_9$.}\label{e8}
\end{figure}

Below we drop the dependence of $S_9(z)$ on $z$ for simplicity. The changes of variables from the ninth charts $(u_{9j1}, u_{9j2})$ ($j=1, 2$) to the original variables shows explicitly that a zero $\zeta$ of $u_{912}$ is a pole of the solution of $\Po$. Moreover, the blow-up of $b_8$ causes $E\,w_{92}$ to be infinite along ${\mathcal E}_{9}\backslash {\mathcal E}_8^{(1)}$, the line determined by $u_{921}=0$. In turn, this means the energy function $E(z)$ has a pole at $z=\zeta$, where the solution $(u_1(z), u_2(z))$ of the Boutroux-Painlev\'e equation has a pole. If $u_{911}(\zeta)=a$, the estimates for large $|z|$ show that $E(z)$ is well approximated by $2^{-7}a$ as soon as $z$ leaves the disc centred at $z=\zeta$ with radius of order $1/|\zeta|$. 

Since the vector field has no base points in $S_9$, is infinite along the configuration 
\[
 I(z):=\bigcup_{i=0}^8\, {\mathcal E}_i^{(9-i)}(z)
\]
and is regular elsewhere in $S_9$, we call the set $I(z)$ the \textit{infinity set} of the vector field. Let ${\mathcal S}$ denote the fiber bundle of the surfaces 
$S_9=S_9(z)$, $z\in\complex\setminus\{ 0\}$, in which the 
time\--dependent Painlev\'e vector field $v_z$, in the Boutroux 
scaling, defines a regular (i.e., holomorphic) 
one\--dimensional vector subbundle ${\mathcal P}$ of ${\mathcal S}$. 
For each $z\in\complex\setminus\{ 0\}$,  the infinity set  
$I(z)$ 
is where ${\mathcal P}$ 
is ``vertical'' (or  tangent to the fiber). 
If ${\mathcal I}$ denotes the union in ${\mathcal S}$ 
of all $I(z)$, $z\in\complex\setminus\{ 0\}$, then 
${\mathcal S}\setminus {\mathcal I}$ is Okamoto's \lq\lq space  
of initial conditions\rq\rq , fibered by the surfaces 
$S_9(z)\setminus I(z)$, the open subset of 
${\mathcal S}$ of all points in ${\mathcal S}$ where ${\mathcal P}$ 
is transversal to the fibers, and therefore defines a 
regular infinitesimal connection in the bundle 
of the $S_9(z)\setminus I(z)$, $z\in\complex\setminus\{ 0\}$. 

The explicit construction of a good resolution of the Painlev\'e-Boutroux system allow us to show that, for every $\epsilon>0$, there exists a neighbourhood $U$ of the union of the first seven exceptional lines 
$
I^6:=\bigcup_{i=0}^6\, {\mathcal E}_i^{(9-i)},
$
in $S_9$ such that $|(\dot{E}/E)/(-6/5\, z)-1|<\epsilon$ in $U$ and for all 
$z\in\complex\setminus\{ 0\}$. For every compact subset 
$K$ of ${\mathcal E}_7^{(2)}\setminus {\mathcal E}_8^{(1)}$ there exists a 
neighbourhood $V$ of $K$ in $S_9$ and a constant $C>0$ such that 
$|(\dot{E}/E)\, z|\leq C$ in $V$  and for all 
$z\in\complex\setminus\{ 0\}$. Moreover, supposing $z$ is bounded away from zero, we can find a measure of the distance to the infinity set $I$, in the following way. Let $q:=2\, E$. 
There exists a continuous complex valued function $d$ on 
a neighborhood of $I$ in $S_9$ such that 
$d=q^{-1}$ in a neighborhood in $S_9$ of $I\setminus 
{\mathcal E}_8^{(1)}$, 
$d=w_{92}$ in a neighborhood in $S_9$ of the remaining part 
${\mathcal E}_8^{(1)}\setminus {\mathcal E}_7^{(2)}$ of $I$, 
and $q\, d\to 1$, $d/w_{92}\to 1$ when approaching 
${\mathcal E}_8^{(1)}\setminus {\mathcal E}_7^{(2)}$. These estimates allow us to prove Result (i) in Remark \ref{rep} below.

To investigate whether our asymptotic description is complete, we define and study a complex version of the limit set of solutions, whose real version is well known in dynamical systems theory. 
\begin{definition}\label{limit set}
For every solution $\complex\setminus\{ 0\}
\ni z\mapsto U(z)\in S_9(z)\setminus I(z)$, 
let $\Omega _U$ 
denote the set of all $s\in S_9(\infty )
\setminus I(\infty )$ such that there 
exists a sequence $z_j\in\complex$ with the property that 
$z_j\to\infty$ and $U(z_j)\to s$ as $j\to\infty$. 
The subset $\Omega _U$ of $S_9(\infty )
\setminus I(\infty )$ is called the 
{\em limit set of the solution $U$}. 
\end{definition}
Using the compactness of $S_9$, and estimates of the vector fields away from $I$, we can prove Result (ii) in Remark \ref{rep} below. Combining Results (i) and (ii) in Remark \ref{rep}, we find Result (iii), which surprisingly appears not to have been proved before. 
\begin{remark}\label{rep} 
\begin{list}{{\bf (\roman{enumi})}}
                        {\usecounter{enumi}
                         \setlength{\parsep}{8pt}
                        \setlength\leftmargin{4pt}
                        }
\item The infinity set $I$ of the vector field is repelling, i.e., every solution which starts in Okamoto's space $S_9\setminus I$ remains 
there for all complex nonzero times. See \cite[Corollary 4.3]{dj:11} for the proof.
\item  For every solution $U$ the limit set 
$\Omega _U$  is invariant under the transformation $T$ 
of $S_9(\infty )$ which in the coordinate system 
$(u_1,\, u_2)$ is given by 
$(u_1,\, u_2)\mapsto (-u_1,\,i\,u_2)$, when 
$q\mapsto -q$ and $E\mapsto -E$. Moreover, $\Omega _U$ is a 
non\--empty, compact and connected subset of 
${\mathcal S}_9$. See \cite[Corollary 4.6]{dj:11} for a proof.
\item  Every solution of the first Painlev\'e equation has  infinitely many poles. See \cite[Corollary 4.7]{dj:11} for a proof.
\end{list}
\end{remark}

\section{The $q$-discrete first Painlev\'e equation in the limit $t\to\infty$}\label{qp1}
The construction of the initial-value space of a difference equation such as q$\Po$, is simpler to describe in $\proj^1\times \proj^1$. This identifies the affine coordinates $(w_{1}, w_{2})$ with homogeneous coordinates as follows
\begin{eqnarray*}
 ([1:w_{1}], [1:w_{2}])&=&([w_1^{-1}:1], [1:w_2])\\
 &=&([1:w_{1}], [w_{2}^{-1}:1])\\
&=& ([w_1^{-1}:1], [w_{2}^{-1}:1])
 \end{eqnarray*}
 Rewritten as a system for $w_1=w$, $w_2=\underline w$, q$\Po$ becomes
 \begin{subequations}\label{qp1 sys}
 \begin{align}
 \overline w_1&=\frac{t\,w_1-1}{t\,w_1^2\,w_2}\\
 \overline w_2&=w_1
 \end{align}
 \end{subequations}
 This system has to be considered simultaneously with its backward iteration. The right side becomes undefined (i.e., $0/0$) at $(w_1, w_2)=(1/t,0)$, which is a base point $b_0$ of this system. Similarly, there is a base point $b_1=(0, q/t)$ which arises from the backward iteration. The full investigation of this system as $t\to\infty$ will be carried out in a separate paper. Here, we point out some new behaviours that occur near these base points as $t\to\infty$.

We investigated the solutions of q$\Po$ that remain stationary to leading order, as $t\to\infty$, which is equivalent to $\overline w\sim w$ and $\underline w\sim w$ as $t\to\infty$. To leading order, stationarity in q$\Po$ (Equation \eqref{eq: qp1 scaled}) implies that $w(w^3-1)={\mathcal O}(1/t)$. Therefore, we obtain either $w=\omega+{\mathcal O}(1/t)$, where $\omega^3=1$, or $w={\mathcal O}(1/t)$. While these two cases appear to be superficially similar, they are distinguished by the fact that $\omega$ is a singularity of the invariant (see Equation \eqref{ap1}) of the leading-order equation. This singular leading order behaviour is far away from base points $b_0$ and $b_1$, as $t\to\infty$, while the solution which satisfies  $w={\mathcal O}(1/t)$ is close to these base points.  Notice that while the flow around a base point could have been resolved by standard techniques from algebraic geometry, in this case, two base points (one from the forward iteration and one from backward iteration) coalesce in the asymptotic limit, creating a more complicated flow near the origin.

For the more difficult case $w\ll 1$ as $t\to\infty$, we obtain the following results \cite{j:13}:
\begin{enumerate}
\item Equation \eqref{eq: qp1 scaled} has a formal power series solution
\begin{equation}\label{eq: formal z}
w(\xi)=\sum_{n=1}^\infty\frac{b_n}{t^n},
\end{equation}
where $b_1=1$, $b_2=0$, $b_3=0$ and
\begin{align}
&b_n=\sum_{r=2}^{n-2}\sum_{k=1}^{r-1}\sum_{m=1}^{n-r-1}\,b_k\,b_{r-k}\,b_m\,b_{n-r-m} q^{(r-2\,k)},\ n\ge 4 \label{z formal rec n }
\end{align}
\item
$b_{3\,p+2}=0$ and $b_{3\,p+3}=0$ for all integers $p\ge 0$.
\item For $|q|>e^2$, the solution of the recurrence relation \eqref{z formal rec n } has the asymptotic behaviour 
\begin{equation}\label{div b}
b_{3\,p+1}\underset{p\to\infty}{=}{\mathcal O}\left(|q|^{3\,p\,(p-1)/2}\,\prod_{k=0}^{p-1}\bigl(1+q^{-3k}\bigr)^2\right).
\end{equation}
\item There exists a true solution, which is asymptotic to the asymptotic series \eqref{eq: formal z} as $t\to\infty$ in a region ${\mathcal T}=t_0\,q^{-\,(n_0+\real^+)}$ (satisfying constraints as described in Remark 3.5 and Lemma 3.8 of \cite{j:13}).
\item This solution is unstable in initial value space.
\end{enumerate}
Notice that Equation \eqref{div b} implies that the series \eqref{eq: formal z} is divergent. In the terminology of \cite{z:1998}, which is concerned with linear $q$-difference equations, the formal series \eqref{eq: formal z} lies in the space of $q$-Gevrey series of order $3$.  Nevertheless, for initial values on each sufficiently large smooth path in $\mathcal S$, there exists a true vanishing solution, unstable in initial-value space, which is asymptotic to the formal power series. Its instability, and existence near a coalescing pair of base points, led us to allocate the solution a distinguished name, i.e. {\it quicksilver} solution, in the sense of being quickly changeable and difficult to hold or contain.

\section{Discussion}\label{dis}
This paper provided an overview of how to carry out asymptotic analysis in the geometric space of initial values associated with the Painlev\'e equations. We also sketched an extension of this viewpoint to discrete Painlev\'e equations. For the Painlev\'e equations, our method filled a gap in conventional asymptotic analysis, enabling poles to be included in the asymptotic description of the solutions. It also provided some new results, such as the infinity of number of poles of general solutions in the complex plane and completeness of the limiting behaviours, which had not been considered before.  For the discrete Painlev\'e equations, it led to asymptotic  behaviours that were not anticipated from those of the corresponding Painlev\'e equations, such as genus-zero leading order behaviours and quicksilver solutions. We hope that this approach will provide further useful analytic information about generic solutions of discrete Painlev\'e equations.

\end{document}